\documentclass[twocolumn]{pasj00}
\usepackage[varg]{txfonts}

\begin{document}
\Received{$\langle$reception date$\rangle$}
\Accepted{$\langle$accepted date$\rangle$}
\Published{$\langle$publication date$\rangle$}
\SetRunningHead{}{}
\title{
Systematic X-ray Analysis of  Radio Relic Clusters with $SUZAKU$
\thanks{Last update: \today}}
\author{
Hiroki Akamatsu\altaffilmark{1} 
and 
Hajime Kawahara\altaffilmark{2}
}
\altaffiltext{1}{
SRON Netherlands Institute for Space Research, Sorbonnelaan 2, 3584 CA Utrecht, The Netherlands
}
\altaffiltext{2}{
Department of Physics, Tokyo Metropolitan University,\\
 1-1 Minami-Osawa, Hachioji, Tokyo 192-0397}
\maketitle

\begin{abstract}
We perform a systematic X-ray analysis of six giant radio relics in four clusters of galaxies using the Suzaku satellite. The sample includes CIZA 2242.8-5301, Zwcl 2341.1-0000, the South-East part of Abell 3667 and previously published results of the North-West part of Abell 3667 and Abell 3376.
Especially we first observed the narrow (50 kpc) relic of CIZA 2242.8-5301 by Suzaku satellite, 
which enable us to reduce the projection effect.
We report X-ray detections of shocks at the position of the relics in CIZA2242.8-5301 and Abell 3667 SE.
At the position of the two relics in ZWCL2341.1-0000, we do not detect shocks. 
From the spectroscopic temperature profiles across the relic, we find that the temperature profiles exhibit significant jumps across the relics for CIZA 2242.8-5301, Abell 3376, Abell 3667NW, and Abell 3667SE.  
We estimated the Mach number from the X-ray temperature or pressure profile using the Rankine-Hugoniot jump condition and compared it with the Mach number derived from the radio spectral index. The resulting Mach numbers (${\cal M}=~$1.5-3) are almost consistent with each other, while the Mach number of CIZA2242 derived from the X-ray data tends to be lower than that of the radio observation. 
These results indicate that the giant radio relics in merging clusters are related to the shock structure, as suggested by previous studies of individual clusters. 
\end{abstract}
\section{Introduction \label{sec:intro}}
The formation theory of galaxy clusters and groups is now well established: the high density peaks that exist within the primordial matter distribution later grow to groups and clusters by both accretion of surrounding matter and mergers of clumps. In the growth history of haloes, major mergers have a significant effect on the internal structure of clusters. X-ray observations have revealed a lot about the merging phenomena of haloes from the massive clusters scale (e.g. \cite{2002ApJ...567L..27M}) to galaxy groups (e.g. \cite{2011ApJ...727L..38K}) mainly by studying X-ray morphology. Shock structures induced by mergers were clearly found in the Bullet cluster using {\it Chandra} \citep{2002ApJ...567L..27M}. In general, however, an identification of a shock structure in X-ray images is difficult, except in major mergers including 1E0657-56, A520, A2146 and A754 (\cite{clowe06,markevitch05,russell10,macario11}). 

Recent progress in radio astronomy has revealed interesting arc structures in merging clusters, called radio relics \citep{ferrari08}. Radio relics are believed to be a tracer of shock structures through synchrotron radiation. \citet{finoguenov10} used XMM-Newton data of the well-known radio relic cluster Abell 3667 and found a sharp temperature decrement across the radio relic. \citet{akamatsu11_a} confirmed their result using data from the Suzaku satellite.  

Here, we present a systematic X-ray analysis of 6 radio relics in 4 clusters observed using Suzaku, including a new observation of a strong radio relic cluster, CIZA 2242.8-5301 (hereafter CIZA 2242), which was recently discovered in a Giant Metrewave Radio Telescope (GMRT) observation \citep{swarup91}. New analyses of archival data of Zwcl 2341.1-0000 (hereafter Zwcl2341), Abell3667SE, and a compilation of previous results of Abell3667NE \citep{akamatsu11_a} and Abell3376 \citep{akamatsu11_b} are also discussed.

We use $H_0 = 70$ km s$^{-1}$ Mpc$^{-1}$, $\Omega_{\rm M}=0.27$ and $\Omega_\Lambda = 0.73$, respectively. We use solar abundances as given by \citet{anders89}. 
Unless otherwise stated, the errors correspond to the 68\% confidence level for a single parameter.
The rest of this paper is organized as follows. In \S 2, we describe the target selection, data reduction, and the X-ray spectral analysis. In \S 3 we derive the temperature across the relic to estimate the Mach number. In \S 4 we summarize the results.

\begin{table*}[t] 
\begin{center}
\small
\caption{Basic properties of the clusters }
\label{tab:target}
\begin{tabular}{cccccccc}  \hline  
	&	redshift		&physical length for $1 \timeform{'}$	&Temperature	& $N_{\rm H}$		& Radio references & \\
	&				&	(kpc)	&	(keV)			&	(10$^{20}$~cm$^{-2}$)&\\ \hline
CIZA2242	& 0.192	&	190	&8.4	& 33.4	&\cite{vanweeren10}\\
Abell3376$\ast$	& 0.046	&	54	&4.2	&5.8		&\cite{bagchi06}, \cite{kale12}\\
Abell3667NW$\dagger$	& 0.056	&	66	&6.0	&4.7		&\cite{rottgering97}		&	\\
Abell3667SE	& 0.056	&	66	&6.0	&4.7		&\cite{rottgering97} 		\\
Zwcl2341N	&0.270	&	246	&5.7	&3.4		&\cite{bagchi02,vanweeren09}	\\ 
Zwcl2341S	&0.270	&	246	&5.7	&3.4		&\cite{bagchi02,vanweeren09}	\\ 
\hline
\multicolumn{6}{l}{$\ast$:\cite{akamatsu11_a}}\\
\multicolumn{6}{l}{$\dagger$:\cite{akamatsu11_b}}\\
\end{tabular}
\end{center}
\end{table*}

\begin{table*}[t]
\begin{center}
\caption{Observation log and exposure time after data screening without and with COR-cut(COR2 $>$ 6GV) }
\label{tab:log}
\begin{tabular}{cccccccc}  \hline  
Name (ObsID)					&(R.A., DEC )		 & Observation Start time & Exp time$^{\dagger}$ \\ \hline 
CIZA2242(806001010)			& (340.74, +53.16) 	& 2011-07-28			& 102.1  \\
CIZA2242 OFFSET(806002010) 	& (339.29, +52.67) 	& 2011-07-30			& 52.4   \\
Abell3667SE(805036010) 		& (303.44, -57.03) 	& 2010-04-12			& 45.0\\
Zwcl2341(803001010) 			& (355.90, +00.34)	& 2008-06-27  		& 41.0 \\
\hline 
$\dagger$:COR2  $>$ 6GV \\
\end{tabular}
\end{center}
\end{table*}

\begin{figure*}[]
\begin{tabular}{cc}
\begin{minipage}{0.45\hsize}
(a)CIZA2242
\begin{center}
\includegraphics[width=0.75\hsize]{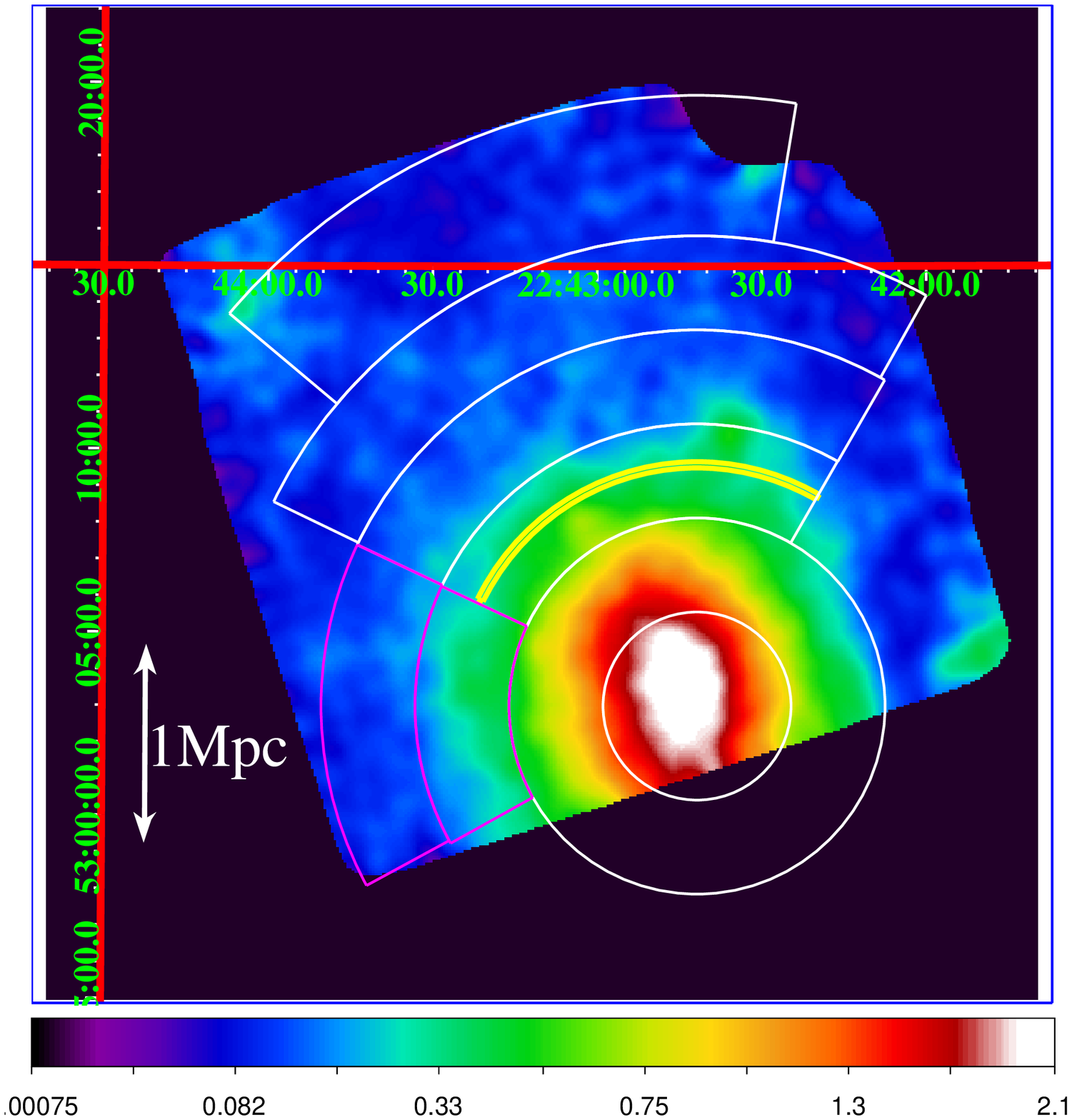}
\end{center}
\end{minipage}
\begin{minipage}{0.55\hsize}
(b)A3376
\begin{center}
\includegraphics[scale=0.45]{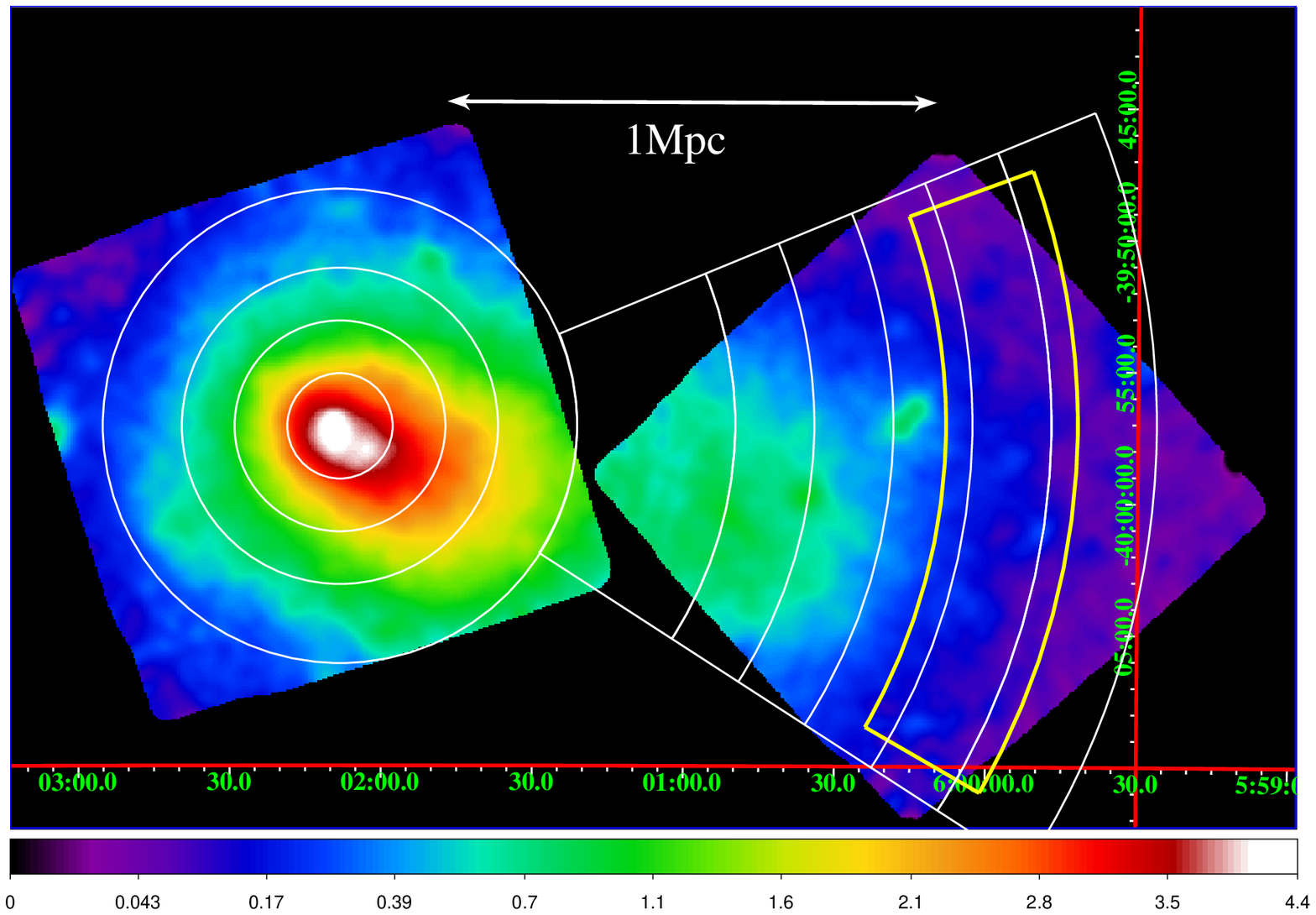}
\end{center}
\end{minipage}\\
\begin{minipage}{0.5\hsize}
(c)A3667
\begin{center}
\includegraphics[width=0.75\hsize]{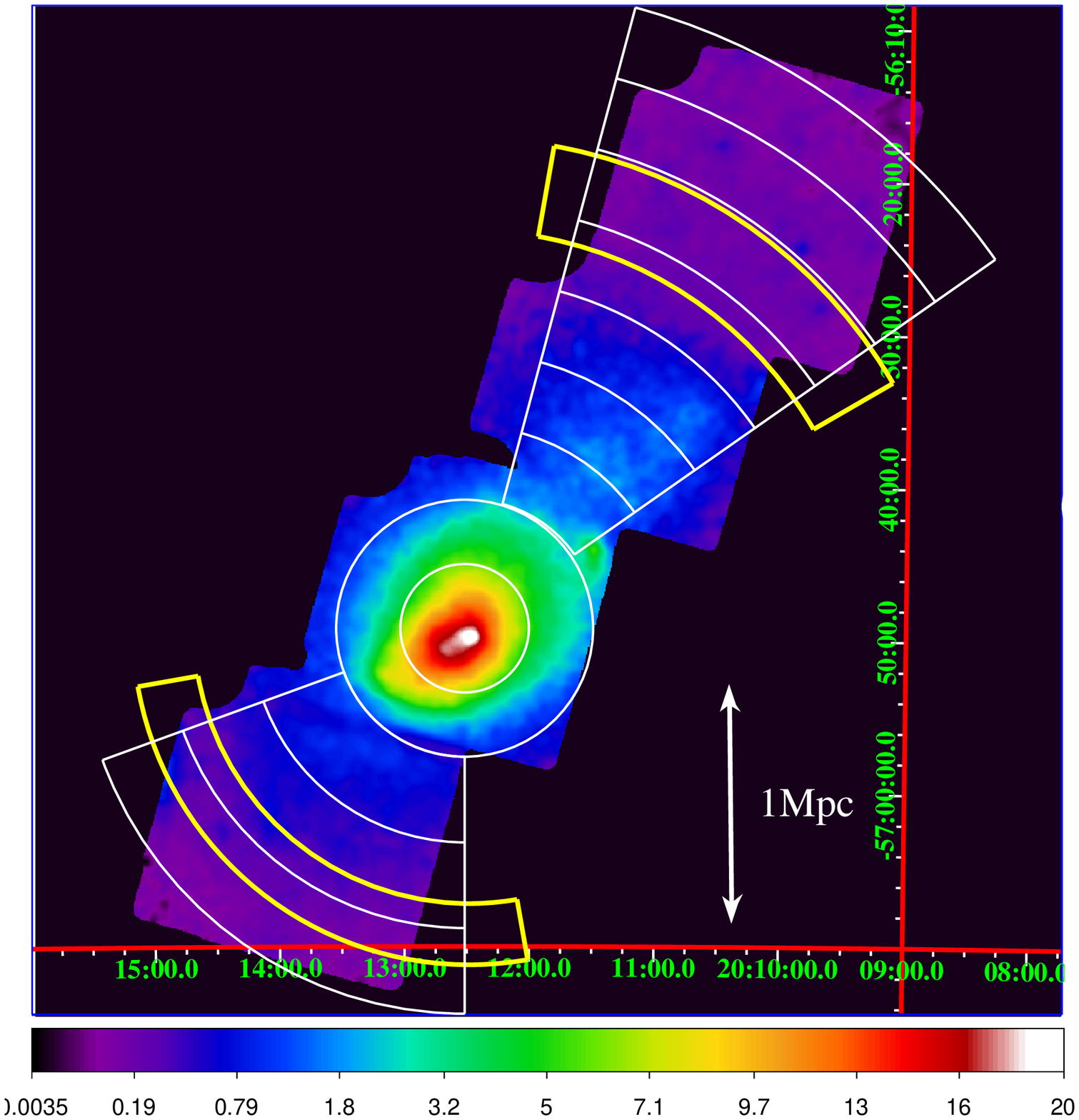}
\end{center}
\end{minipage}
\begin{minipage}{0.5\hsize}
(d)Zwcl2341
\begin{center}
\includegraphics[width=0.75\hsize]{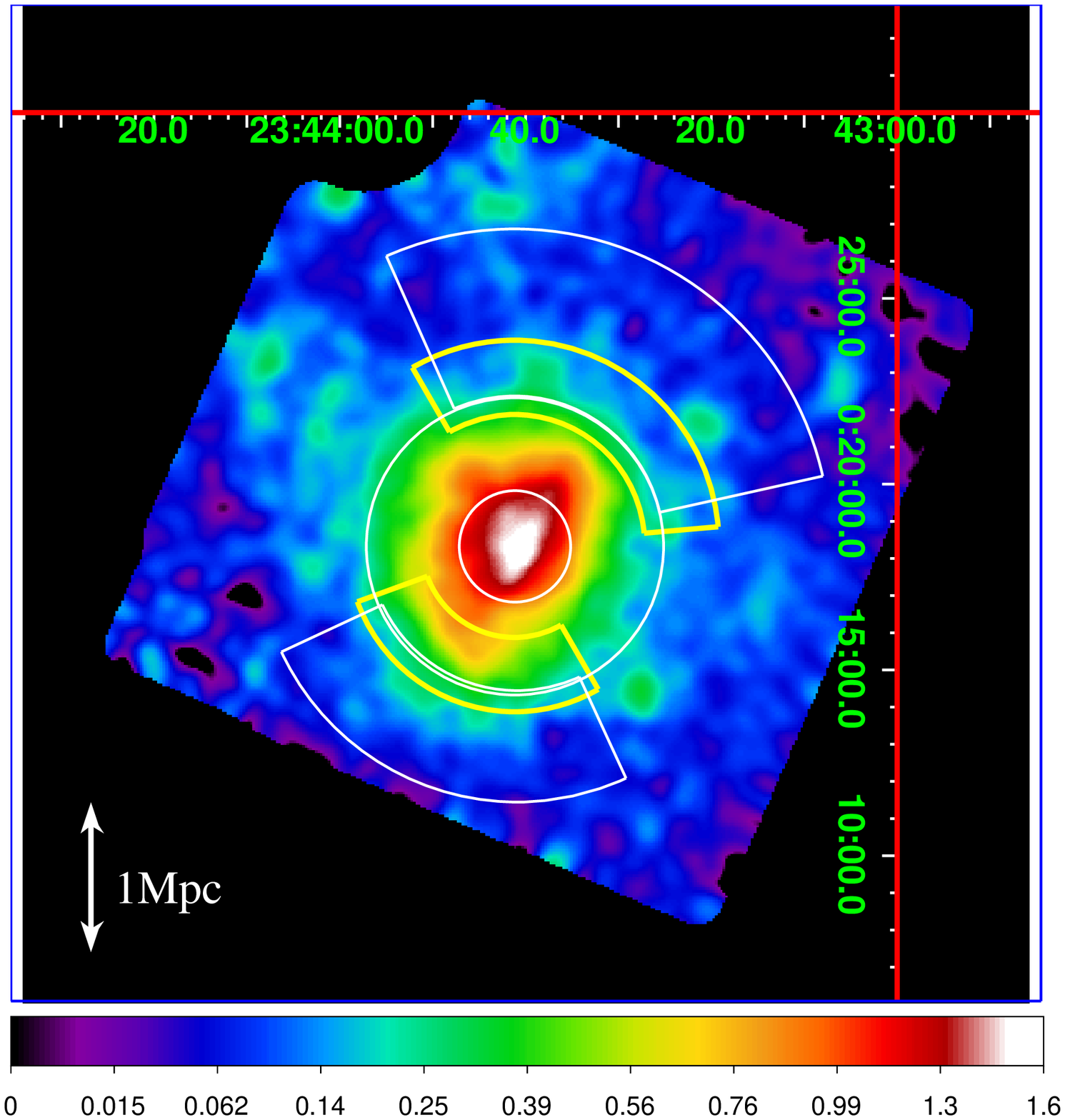}
\end{center}
\end{minipage}\\
\end{tabular}
\caption{
X-ray image of (a)CIZA2242, (b)A3376, (c)A3667 and (d)Zwcl2341 in the energy band 0.5-8.0 keV, after
  subtraction of the NXB without vignetting correction and after
  smoothing by a 2-dimensional Gaussian with $\sigma =16$ pixel
  =$\timeform{17''}$. White and Yellow annuli indicate the spectrum analysis regions and the radio relic region. 
 Magenta annuli in Panel (a) is used for the analysis of the region perpendicular to the merger axis.}
\label{fig:image}
\end{figure*}

\section{Observations \& Data preparation}

\subsection{Targets}
In Table~\ref{tab:bgd}, we summarize the information about the relics and clusters we analyze. Four relics (CIZA2242, Abell3667SE, Zwcl2341N, and Zwcl2341S) are newly analyzed in this paper and our previous analysis are used for two relics (Abell3667NW, A3376). We examine these clusters one by one.

CIZA2242  ($z=0.192$) was discovered in the Clusters in the Zone of Avoidance (CIZA) survey~\citep{kocevski07}.  \citet{vanweeren10} reported a Mpc scale radio relic located in the Northern outskirts, at a distance of 1.2 Mpc from the cluster center. The radio relic is extremely narrow with a width of 55 kpc. The spectral index at the front of relic is -0.6 $\pm$ 0.05, which corresponds to a Mach number of ${\cal M}=4.5^{+1.3}_{-0.9}$. Although this shock wave is the strongest of all the shocks in this sample, only X-ray data from ROSAT have been available so far. We obtained a 120 ksec Suzaku observation of CIZA2242 performed on September 28, 2011 with an additional 60 ksec offset observation.  

A3376 ($z=0.046$) is considered to be experiencing a binary subcluster merger. Previous BeppoSAX/PDS observations showed a 2.7 $\sigma$ detection of a hard X-ray signal~\cite{nevalainen04}. However, Suzaku HXD observations gave an upper limit, which did not exclude the BeppoSAX flux \citep{kawano09}. The detection of a hard X-ray signal is still controversial. 
Another striking feature of A3376 is a pair of Mpc-scale radio relics~\citep{bagchi06, kale12}.  
Using Suzaku data, \citet{akamatsu11_b} confirmed evidence for the presence of shock fronts across the west radio relic.

A3667 ($z = 0.0556$) is well known as a merging cluster with irregular X-ray morphology 
and has large two extended radio relics at the North-West (A3667NW) and the South-East (A3667SE) \citep{rottgering97}. The overall spectral index of North-West relic gradually varies -1.1 to -1.5 toward the southeast direction. A3667 exhibits an elongated X-ray shape to the North-West direction with an average temperature of $7.0 \pm 0.5$ keV \citep{knopp96, briel04}. Recent XMM-Newton and Suzaku observations of the North-West radio relic revealed a significant jump in temperature and surface brightness across the relic~\citep{finoguenov10,akamatsu11_a}. The observed ICM properties indicate the presence of a shock front with Mach number ${\cal M} \sim 2.4$. The South-East radio relic has not yet been studied in X-rays.

Zwcl2341 ($z=0.270$) has two Mpc-scale structures of diffuse radio emission, which were discovered in 1.4 GHz New VLA All the Sky Survey (NVSS)~\citep{bagchi02} at the North~(Zwcl2341N) and South~(Zwcl2341S) of the clusters. These radio relics are located at a Northern (Zwcl2341N $\sim$860 kpc ) and Southern (Zwcl2341S$\sim$ 1100 kpc) 
direction from the cluster center ~\citep{bagchi02}. Recently, GMRT observations revealed that those radio emissions exhibit "arc"-like shapes~\citep{vanweeren09}. The spectral indices of the relics are -0.49$\pm$0.18 for Zwcl2341N and -0.76$\pm$0.17 for Zwcl2341S corresponding to Mach numbers ${\cal M} > 3.57$ and $2.95 \pm 1.39$, respectively.
Previous Chandra and XMM-Newton observations~\citep{vanweeren09} show that 
the X-ray emission extends over $\sim$ 3.3 Mpc in the North-South direction. Due to its low X-ray surface brightness, the X-ray information around these radio relics is still limited.

\subsection{Data reduction}
All SUZAKU observations of CIZA2242, CIZA2242 OFFSET, Abell3667SE and Zwcl2341 were performed with the XIS in either normal $5\times5$ or $3\times3$ clocking mode. The detailed information about these observations is summarized in Table~\ref{tab:log}. For all data, 3 out of the 4 CCD chips were available: XIS0, XIS1 and XIS3. The XIS1 is the back-illuminated chip with high sensitivity in the soft X-ray energy range.  

We use HEAsoft version 6.11 and the calibration database (CALDB) dated 2011-06-30. We perform the event screening with the cut-off rigidity (COR) $> 6$ GV to increase the signal-to-noise ratio. For CIZA2242, we set the Earth rim ELEVATION $> 10^{\circ}$ to avoid contamination by scattered solar X-rays from the day Earth limb. 
Furthermore, we apply additional processing for XIS1 to reduce NXB level, which
increased after changing the amount of charge injection (06/2010).
The detailed processing procedure is the same as the one described in $XIS~analysis~topics$
\footnote{http://www.astro.isas.jaxa.jp/suzaku/analysis/xis/xis1\_ci\_6\_nxb/}.
Regions affected by the calibration sources were masked out using the {\it calmask} CALDB file. 
We used data in the energy range between 0.5--10~keV for the FI detectors and between 0.5--8~keV for the BI detector. 
For CIZA2242, we used the energy range between 0.75--10~keV for the FI detectors and 0.6--8~keV for the BI detector.

\subsection{X-ray Spectrum Analysis}
In this paper, we follow the same spectral fitting procedure as for the Suzaku observations of A3667 and A3376 ~\citep{akamatsu11_a, akamatsu11_b}. The basic method of the spectral fit is described in subsection 4.1 of \citet{akamatsu11_a}. 

\subsubsection{Background and Foreground Emissions}\label{sec:bgd}
An accurate measurement of the background components is crucial for the
ICM study in the cluster outskirts. We consider three sky background components: the local hot bubble (LHB), the Milky way halo (MWH), and the cosmic X-ray background (CXB). We analysed the offset region for CIZA2242OFFSET (ID:806002010), which is one degree away from the main target. For the A3667SE observation, we used the background model reported in \cite{akamatsu11_a} Table3 and for Zwcl2341, we used the outer region where there is negligible ICM emission.

The non X-ray background (NXB) subtracted spectra were fitted by the summation of the LHB: ($\sim$0.1 keV), the MWH ($\sim$0.3 keV) and the CXB as  ${\it apec+wabs(apec+powerlaw)}$.  We fix the redshift and abundance of all {\it apec} components to zero and unity, respectively. 
The NXB component was estimated from the dark Earth database 
by the ${\it xisnxbgen}$ FTOOLS \citep{tawa08} and
was subtracted from the data before the spectral fit. 
We used the XIS response assuming uniform brightness on the sky by {\it xissimarfgen} \citep{ishisaki07}. 
The best fit parameters and spectrum of CIZA2242OFFSET are shown in Table~\ref{tab:bgd} and Fig.~\ref{fig:bgd}, respectively.

\begin{figure*}[htbp]
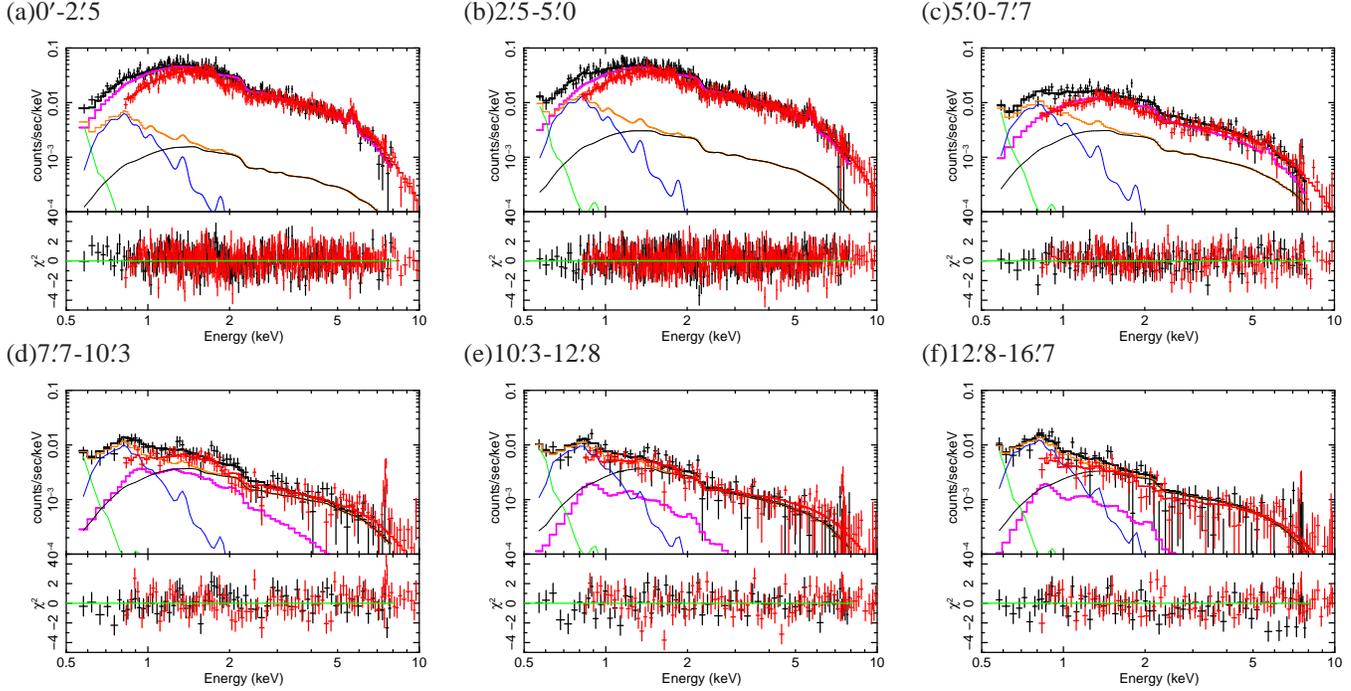

\begin{tabular}{cc}
\begin{minipage}{0.333\hsize}
(a)\timeform{0'}-\timeform{2.5'}
\\[-0.8cm]
\begin{center}
\includegraphics[angle=-90,scale=0.22]{fig/ICM-check-ciza-1-cxb+.ps}
\end{center}
\end{minipage}
\begin{minipage}{0.3333\hsize}
(b)\timeform{2.5'}-\timeform{5.0'} 
\\[-0.8cm]
\begin{center}
\includegraphics[angle=-90,scale=0.22]{fig/ICM-check-ciza-2-cxb+.ps}
\end{center}
\end{minipage}
\begin{minipage}{0.3333\hsize}
(c)\timeform{5.0'}-\timeform{7.7'} 
\\[-0.8cm]
\begin{center}
\includegraphics[angle=-90,scale=0.22]{fig/ICM-check-ciza-3-cxb+.ps}
\end{center}
\end{minipage}\\
\begin{minipage}{0.3333\hsize}
(d)\timeform{7.7'}-\timeform{10.3'} 
\\[-0.8cm]
\begin{center}
\includegraphics[angle=-90,scale=0.22]{fig/ICM-check-ciza-4-cxb+.ps}
\end{center}
\end{minipage}
\begin{minipage}{0.3333\hsize}
(e)\timeform{10.3'}-\timeform{12.8'} 
\\[-0.8cm]
\begin{center}
\includegraphics[angle=-90,scale=0.22]{fig/ICM-check-ciza-5-cxb+.ps}
\end{center}
\end{minipage}
\begin{minipage}{0.33333\hsize}
(f)\timeform{12.8'}-\timeform{16.7'} 
\\[-0.8cm]
\begin{center}
\includegraphics[angle=-90,scale=0.22]{fig/ICM-check-ciza-6-cxb+.ps}
\end{center}
\end{minipage}\\
\end{tabular}
\caption{
The NXB subtracted spectra of six annuli of the CIZA2242 (corresponding to white annuli in Figure \ref{fig:image}.  
The XIS BI (Black) and FI (Red) spectra are fitted with the ICM model ({\it wabs + apec}), 
along with the sum of the CXB and the Galactic emission ({\it apec + wabs(apec + powerlaw)}).  
The ICM components are shown in magenta.
The CXB component is shown with a black curve, and the LHB and MWH emissions are indicated 
by green and blue curves, respectively.  
The total background components are shown by the orange curve. 
}
\label{fig:ciza_spec}
\end{figure*}

\begin{figure}[t]
\begin{center}
\includegraphics[scale=0.35,angle=-90]{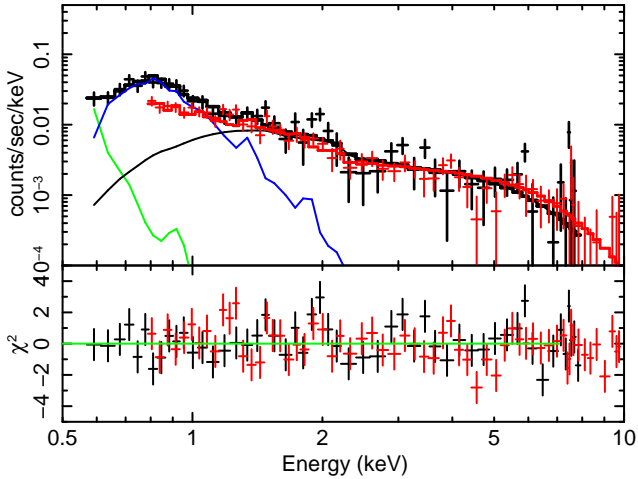}
\end{center}
\caption{ The spectrum of the CIZA2242OFFSET used for the background
  estimation, after subtraction of NXB and the source.  The XIS BI
  (Black) and FI (Red) spectra are fitted with CXB + Galactic
  components (LHB, MWH) ({\it apec+wabs(apec+powerlaw)}).  The CXB
  spectrum is shown with a black curve, and the LHB and MWH components
  are indicated by green and blue curves, respectively.  }
\label{fig:bgd}
\end{figure}

\begin{table}[h]
\begin{center}
\caption{Summary of the best fit parameters for each background observations. \label{tab:cl}}
\begin{tabular}{lccccccc}  \hline  
			&CIZA2242	&A3667SE	&Zwcl2341\\ \hline
LHB&   \\
$kT$(keV)	& 0.08 (fix)	&	0.11$\pm0.01$		&	0.10$\pm0.01$ \\
$norm^{\ast} ~(\times 10^{-3})$  &		12.26	$_{-0.38}^{+0.37}$& 6.85$_{-1.01}^{+1.70}$	&4.85$_{-1.01}^{+1.70}$	\\
\hline
MWH\\
$kT$(keV)	&	0.29$\pm 0.01$		&$0.37\pm0.03$ 	&$0.25\pm0.02$\\
$norm^{\ast}~(\times 10^{-4})$ &	55.70$_{-5.1}^{+5.2}$	&$11.6_{-1.6}^{+1.2}$&	1.1$_{-0.2}^{+0.3}$ \\
\hline
CXB ($\Gamma=1.41$)\\
$norm^{\dagger}~(\times 10^{-4})$ 	&	8.8	$\pm0.4$	&9.6$\pm0.6$		&11.2 $\pm 0.3$\\
\hline
\multicolumn{4}{l}{\footnotesize
*:Normalization of the apec component scaled with a factor 1/400$\pi$.}\\
\multicolumn{4}{l}{\footnotesize
Norm=$\rm \frac{1}{400\pi} \int n_{e}n_{H} dV/(4\pi(1+z^2)D_{A}^2)\times~10 ^{-14} \rm ~cm^{-5}~arcmin^{-2}$, }\\
\multicolumn{4}{l}{\footnotesize
where $D_A$ is the angular diameter distance to the source.}\\
\multicolumn{4}{l}{\footnotesize
The referred CXB intensity normalization in \cite{kushino02} is }\\
\multicolumn{4}{l}{\footnotesize
9.6$~\times10^{-4}$ for $\Gamma=1.41$ in the unit of photons keV$^{-1}~\rm cm^{-2}~s^{-1}$ at 1 keV.}
\end{tabular}
\label{tab:bgd}
\end{center}
\end{table}

\subsubsection{Spectrum fitting}\label{sec:model}
We extracted spectra from regions defined by their distances from the center ($\timeform{22h42m41.9s}, \timeform{50D03m00s}$) {$~\timeform{2.5'}, ~\timeform{5.0'}, ~\timeform{7.7'}, ~\timeform{10.3'}, ~\timeform{12.8'}, ~\timeform{16.7'}$} (Fig~\ref{fig:image}a ) for CIZA2242, 
$~\timeform{16.8'}, ~\timeform{21.0'}, ~\timeform{25.2'}$ centered on  ($\timeform{20h12m31s}, \timeform{-56D49m12s}$) (Fig~\ref{fig:image}b )
 for A3667SE, 
{$~\timeform{1.5'}, ~\timeform{4.0'}, ~\timeform{8.5'}, ~\timeform{10.0'}$ centered on ($\timeform{23h43m41s}, \timeform{+00D18m19s}$) for Zwcl2341N, 
{$~\timeform{1.5'}, ~\timeform{4.0'}, ~\timeform{7.0'}$ centered on ($\timeform{23h43m41s}, \timeform{+00D18m19s}$)
 (Fig~\ref{fig:image}d ) for Zwcl2341S,  respectively.  
For CIZA2242, the center was determined by visually fitting a circle to the radio relics. For other clusters, the center was determined from their X-ray surface brightness peaks.

The response matrix files (RMFs) and ancillary response
files (ARFs) were generated using {\it xisrmfgen} and {\it xissimarfgen}~\citep{ishisaki07}.
Since the Galactic and CXB components can be regarded as being spatially uniform in the XIS field of view, we use an uniform ARF for these components.
For the ICM emission, we need to put the flux distribution into {\it xissimarfgen}.
For A3667 and A3376, we adopted the $\beta$-model distribution with $\beta$=0.54, $r_{c}$=\timeform{2.97'} and 
$\beta=0.40$, $r_{c}=\timeform{2.03'}$, respectively.
For CIZA2242 and Zwcl2341, we used the Suzaku XIS image (0.5-8.0 keV).

The spectral fitting of each region was performed as described below. We modeled the NXB subtracted spectrum as the sum of the ICM emission and each background component discussed in Sec~\ref{sec:bgd}. Intensity, temperature, photon indices of CXB and galactic components were all fixed at the value determined in offset observation. 
The interstellar absorption was kept fixed using the twenty-one centimeter line measurement of the hydrogen column, 
$N_{\rm H}=33.4, ~4.7 \rm ~and ~3.4 \times 10^{20}\rm ~cm^{-2}$ for CIZA2242, A3667SE and Zwcl2341, respectively~\citep{dickey90}. 
We simultaneously fitted both the BI and FI spectra with XSPEC ver12.7. 
For the inner regions of the radio relic, the free parameters were temperature $kT$, normalization {\it norm} and 
metal abundance $Z$ of the ICM component. 
Otherwise, we fixed the ICM metal abundance to the typical value in cluster outskirts, $Z=0.2 Z_\odot$ (e.g. \cite{fujita08}). 
The relative normalization between the two sensors was a free parameter in this fit to compensate for cross-calibration errors.  

\ifnum1=0
\begin{figure}[t]
\begin{center}
\includegraphics[scale=0.35,angle=-90]{fig/ICM-check-ciza-77.ps}
\end{center}
\caption{
The NXB subtracted spectra in CIZA2242.
The XIS BI (Black) and FI (Red) spectra are fitted with the ICM model ({\it wabs + apec}),
along with the sum of the CXB and the Galactic emission ({\it apec + wabs(apec + powerlaw)}).
The CXB component is shown with a black curve, and the LHB and MWH emissions are indicated
by green and blue curves, respectively.
}
\label{fig:ciza_spec}
\end{figure}
\fi

\begin{figure*}[t]
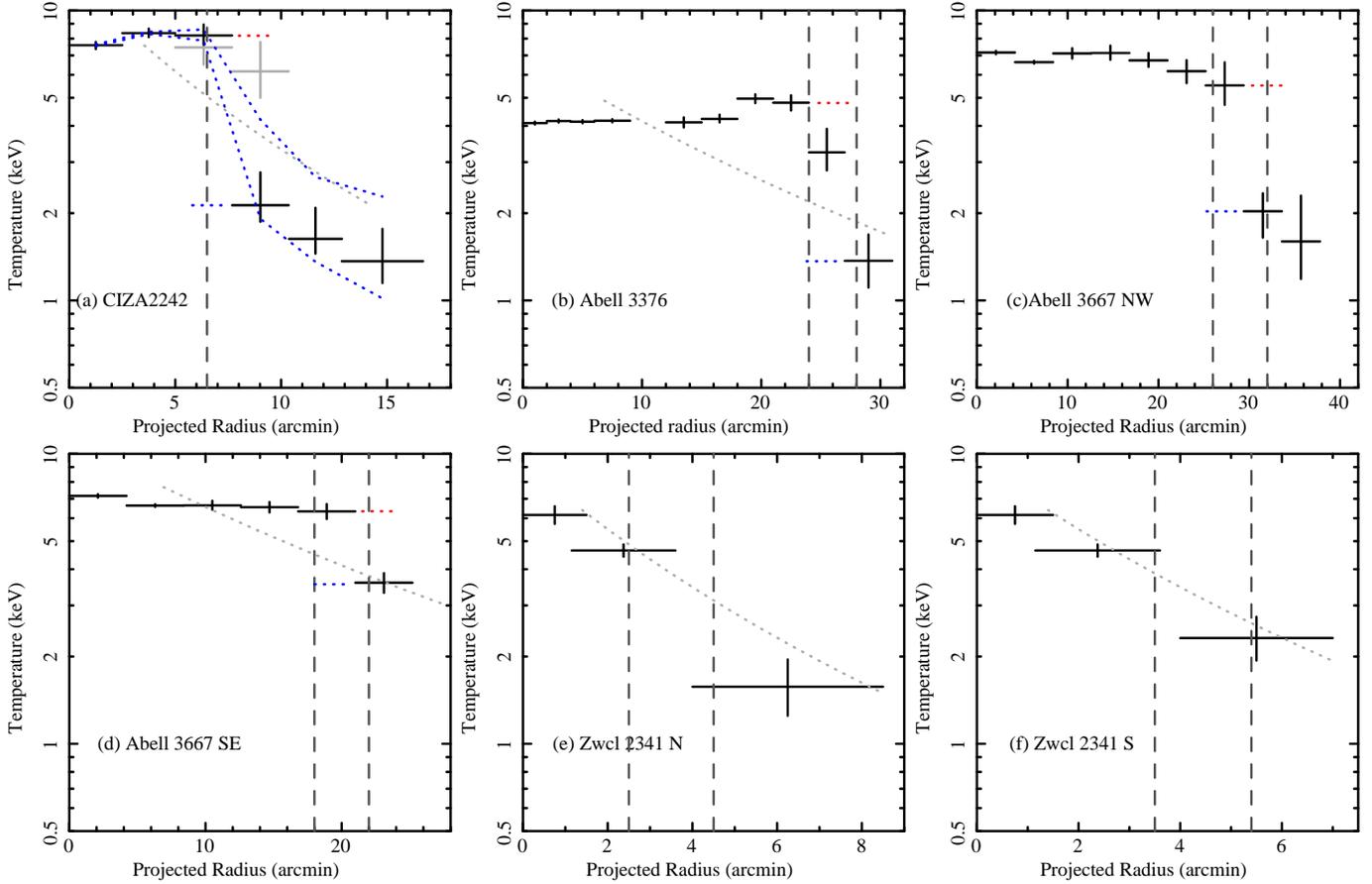

\begin{tabular}{ccc}
\begin{minipage}{0.33333\hsize}
\begin{center}
\includegraphics[angle=-90,width=1.\hsize]{fig/kt_ciza.ps}
\end{center}
\end{minipage}
\begin{minipage}{0.33333\hsize}
\begin{center}
\includegraphics[angle=-90,width=1.\hsize]{fig/kt_a3376.ps}
\end{center}
\end{minipage}
\begin{minipage}{0.33333\hsize}
\begin{center}
\includegraphics[angle=-90,width=1.\hsize]{fig/kt_a3667NE.ps}
\end{center}
\end{minipage}\\
\begin{minipage}{0.33333\hsize}
\begin{center}
\includegraphics[angle=-90,width=1.\hsize]{fig/kt_a3667SE.ps}
\end{center}
\end{minipage}
\begin{minipage}{0.33333\hsize}
\begin{center}
\includegraphics[angle=-90,width=1.\hsize]{fig/kt_zwcl2543N.ps}
\end{center}
\end{minipage}
\begin{minipage}{0.33333\hsize}
\begin{center}
\includegraphics[angle=-90,width=1.\hsize]{fig/kt_zwcl2543S.ps}
\end{center}
\end{minipage}\\
\end{tabular}
\caption{
X-ray temperature profiles for the six radio relics. Vertical dashed lines indicate the position the of radio relics.  Gray dotted lines indicate the "universal" temperature profile expected from the scaled temperature profiles \citep{burns10}. The red and blue horizontal bars show the pre and post shock quantities we use to derive Mach number (Sec~\ref{sec:Mach}). The vertical bars indicate a $1 \sigma$ error. 
In (a)CIZA2242, the gray crosses show 
the data for a reference sector as shown in Fig. 1(a) magenta annuli.
The uncertainty range due to the fluctuations of CXB intensity are shown by two blue dotted lines.
}
\label{fig:kt} 
\end{figure*}

\section{Results}
\subsection{ICM properties of individual cluster} \label{se:profile}

\subsubsection{CIZA2242} \label{sec:ciza}
We first perform the spectral fitting within the radio relic ($\sim$ 1.2 Mpc=$~\timeform{6'5}$): the temperature and the metal abundance are $kT=7.88\pm0.20$ keV and $Z=0.28\pm0.03$, respectively. 
The unabsorbed flux and luminosity within the radio relic in the 0.5$-$10 keV band are $F_{\rm 0.5-10.0~keV}=8.5\times10^{-12} \rm ~erg~s~cm^{-2}$ and $L_{\rm 0.5-10.0~keV}=8.4\times10^{44}\rm~erg~s $, respectively.

Then, we performed the spectral fitting for each annulus examined in Sec~\ref{sec:model}. Figure~\ref{fig:ciza_spec} displays spectra of six annuli. Figure~\ref{fig:kt}(a) shows the radial profile of the ICM temperature in CIZA2242. 
The temperature within $\timeform{6.5'}$ (1.25 Mpc) from the cluster center shows a fairly 
constant value of $\sim 8-9$ keV.  As mentioned in section\ref{sec:intro}, ~\citet{vanweeren10} reported the sharp radio relic at that radius, which is indicated by vertical dashed line in Panel (a). We found that the temperature profile exhibits a significant jump from 8.3 keV to 2.1 keV across the radio relic. 
Because the ICM emission is lower than the background level in the outer bins ($r>7.7'$),
we consider the variability of the CXB as a systematic error.
The fluctuation of CXB intensity was estimated based on our previous work~\citep{akamatsu11}.
The estimated fluctuations span 15-25 \%.
The resultant parameters after taking into account the systematic
error are shown blue dotted lines in figure~\ref{fig:kt} (a).
The temperature jump is consistent with the prediction of numerical simulation by \citet{vanweeren11}. 
For a reference, we plotted the "universal" temperature profile expected from the scaled temperature profiles \citep{burns10}.

As shown in Fig. 5(a), we could not see the surface brightness jump across the relic directly in the profile. However, in
Appendix we show that this is caused by the limited Suzaku spatial
resolution, and the observed profile is in fact consistent with the expected density jump

\subsubsection{A3376}
Previous studies of A3376 found that the global mean temperature is 4.0 keV and showed the presence of a pair of  Mpc-scale radio relics consistent with 1.4 GHz VLA NVSS observations~\citep{bagchi06}. 
A detailed study of the A3376 radio relic has already been published by \citet{akamatsu11_b}.
We refer to that paper for the images and X-ray information and we will include A3376 in the discussion.

\begin{figure*}[t]
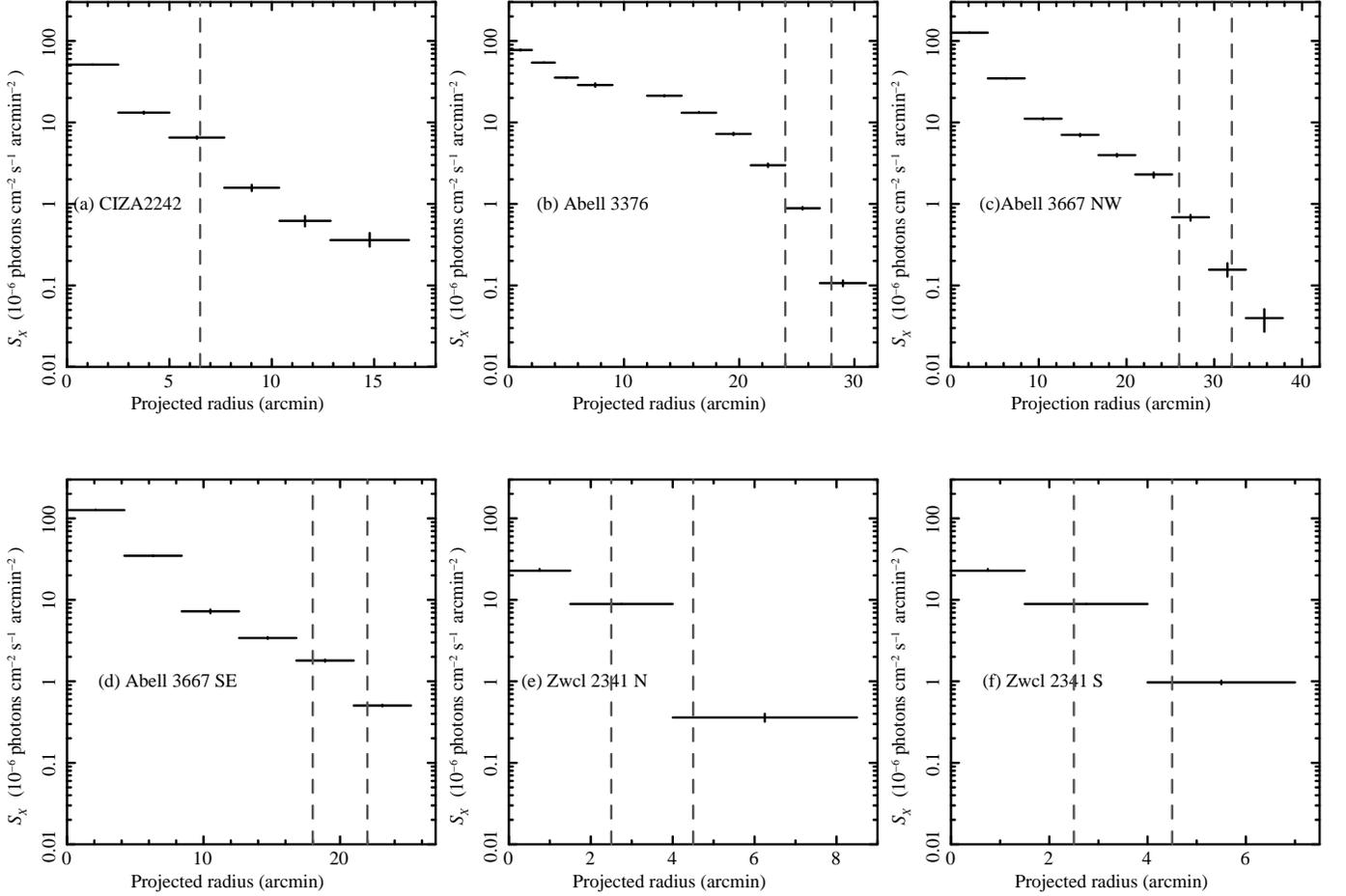

\begin{tabular}{ccc}
\begin{minipage}{0.33333\hsize}
\begin{center}
\includegraphics[angle=-90,width=1.\hsize]{fig/sb_ciza_v2.ps}
\end{center}
\end{minipage}
\begin{minipage}{0.33333\hsize}
\begin{center}
\includegraphics[angle=-90,width=1.\hsize]{fig/sb_a3376.ps}
\end{center}
\end{minipage}
\begin{minipage}{0.33333\hsize}
\begin{center}
\includegraphics[angle=-90,width=1.\hsize]{fig/sb_a3667ne.ps}
\end{center}
\end{minipage}\\
\begin{minipage}{0.33333\hsize}
\begin{center}
\includegraphics[angle=-90,width=1.\hsize]{fig/sb_a3667se.ps}
\end{center}
\end{minipage}
\begin{minipage}{0.33333\hsize}
\begin{center}
\includegraphics[angle=-90,width=1.\hsize]{fig/sb_z2341n.ps}
\end{center}
\end{minipage}
\begin{minipage}{0.33333\hsize}
\begin{center}
\includegraphics[angle=-90,width=1.\hsize]{fig/sb_z2341s.ps}
\end{center}
\end{minipage}\\
\end{tabular}
\caption{
Surface brightness profiles (0.5-10 keV). Vertical dashed lines indicate the position of radio relics. 
\label{fig:sb}
}
\end{figure*}

\subsubsection{Abell3667NW}
Suzaku performed 3 pointing observations of Abell 3667 along the North-West merger axis in May 2006. Figure.~\ref{fig:image}(c) displays the NXB subtracted X-ray image of A3667.
 A detailed study of A3667NW radio relic has already been published by \citet{akamatsu11_a}.
We refer to that paper for the images and X-ray information and we will include A3667NW in the discussion.

\ifnum0=1
The resultant ICM profiles are shown in Fig.~\ref{fig:kt}(c), Fig.~\ref{fig:sb}(c) and Fig.~\ref{fig:pres}(c). The ICM temperature of A3667 gradually decreases toward the outer region from about 7 keV to 5 keV and then shows a drop to 2 keV at the radio relic. The temperature profile inside the relic shows an excess over the mean values for other clusters observed with XMM-Newton~\citep{pratt07}.
We could not see a clear surface brightness drop while \citet{finoguenov10} reported the drop. This might be caused by the low spatial resolution of SUZAKU.
\fi

\subsubsection{A3667SE}
Figure.~\ref{fig:kt}(d) indicates the radial profile of the ICM temperature in the South-East direction of A3667, respectively. The innermost 2 bins of the cluster are identical to that of A3667NW. 
 The temperature within the SE radio relic ($\sim$ 1.3 Mpc) from the cluster center shows a fairly constant value of $\sim 7$ keV,  which is consistent with the NW direction~\citep{akamatsu11_a}. Our temperature profile exhibits a jump across the SE radio relic region. Although we could not see the surface brightness jump (Fig.~\ref{fig:sb}d), the pressure profiles (Fig.~\ref{fig:pres}d) exhibit a slight drop due to the temperature jump.

\subsubsection{Zwcl2341N $\&$ Zwcl2341S}
We first extract the spectrum within 3' ($\sim$ 740 kpc) and fix the redshift value to 0.270. The spectrum was well fitted with the above model ($\chi^{2}_\nu=1.04$ for 473 degrees of freedom). We find a temperature and metal abundance of $kT=5.16\pm0.32$ keV and $Z=0.25\pm0.08$, respectively. The unabsorbed flux and luminosity in the 0.5$-$10 keV band are $F_{\rm 0.5-10.0~keV}=1.0\times10^{-12} \rm ~erg~s~cm^{-2}$ and $L_{\rm 0.5-10.0~keV}=2.2\times10^{44}\rm~erg~s $, respectively. The derived temperature is consistent with previous Chandra observations ($\sim$ 5 keV) \citep{vanweeren09}.

The radial profiles of the ICM temperature are shown in Fig.~\ref{fig:kt}(e) and (f) for Zwcl2341N and Zwcl2341S, respectively.  The temperature drops from 4keV to 1 keV toward the Northern direction or from 4 keV to 2 keV toward the Southern direction.
Figure~\ref{fig:sb}(e,f) and Figure~\ref{fig:pres}(e,f) show the surface brightness and  the pressure profiles of Zwcl2341N and Zwcl2341S. We could not see any jump in our wide radial bins for temperature, surface brightness, and pressure profiles for this cluster.

\begin{figure*}[]
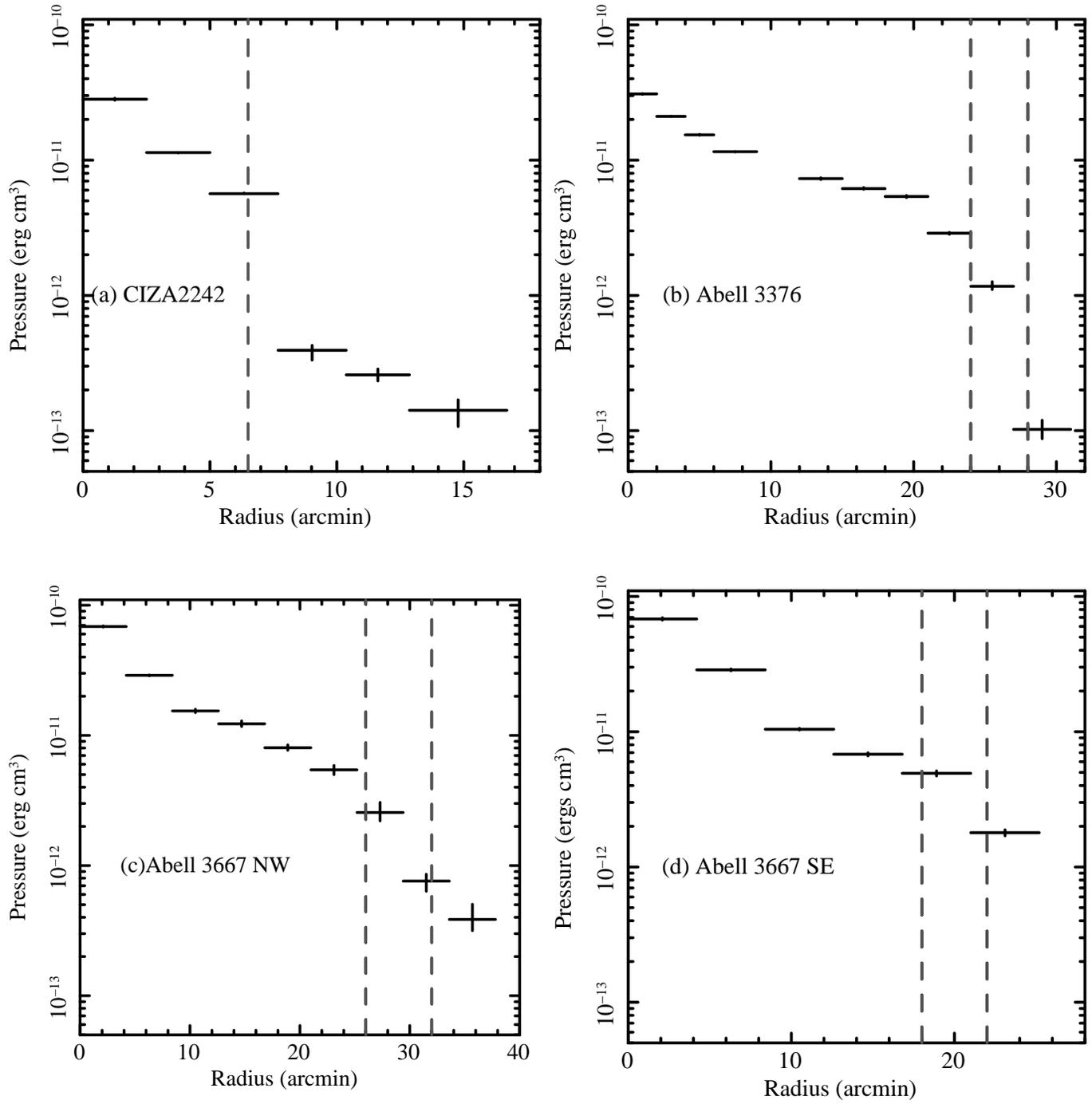

\begin{tabular}{cc}
\begin{minipage}{0.5\hsize}
\begin{center}
\includegraphics[angle=-90,width=1.\hsize]{fig/pres_ciza_v2.ps}
\end{center}
\end{minipage}
\begin{minipage}{0.5\hsize}
\begin{center}
\includegraphics[angle=-90,width=1.\hsize]{fig/pres_a3376.ps}
\end{center}
\end{minipage}\\
\begin{minipage}{0.5\hsize}
\begin{center}
\includegraphics[angle=-90,width=1.\hsize]{fig/pres_a3667nw.ps}
\end{center}
\end{minipage}
\begin{minipage}{0.5\hsize}
\begin{center}
\includegraphics[angle=-90,width=1.\hsize]{fig/pres_a3667se.ps}
\end{center}
\end{minipage}
\end{tabular}
\caption{
Pressure profiles derived by the deprojection technique as described in \citet{akamatsu11}.   Vertical dashed lines indicate the position of radio relics. 
The vertical bars are 1 $\sigma$ errors. 
\label{fig:pres}
}
\end{figure*}

\begin{table*}[t] 
\begin{center}
\caption{The post- and pre- shock ICM quantities and Mach number derived from X-ray and radio observations.}
\label{tab:mach}
\begin{tabular}{cccccccccccc}  \hline  
	&	$kT_{poste~shock}$	&$kT_{pre~shock}$	& ${\cal M}_{X,kT}$	&$\alpha$ &${\cal M}_{\rm radio}$\\
	&		(keV)	&	(keV)	& &		&	\\ \hline
CIZA2242	& 8.33$\pm0.80\pm0.40{\ast}$	& 2.11$\pm0.44 _{-0.20}^{+2.10}{\ast}$	&3.15$\pm0.52_{-1.20}^{+0.40}{\ast}$
&-0.60$\pm$0.05	&4.58$\pm$1.32\\
Abell3376	& 4.81$\pm0.29$	& 1.35$\pm0.35$	&2.94$\pm0.60$	
&	-1.00$\pm$0.10	&2.23$\pm$0.20\\
Abell3667NW	& 5.52$\pm0.95$	& 2.03$\pm 0.34$ 	&2.41$\pm0.39$	
&-1.10$\pm$0.20		&2.08$\pm$0.37\\
Abell3667SE	& 6.34$\pm0.38$	&3.59$\pm0.28$	&1.75$\pm0.13$
&-1.50 $\pm$0.17	&1.73$\pm$0.58\\
\hline
\multicolumn{6}{l}{$\ast$:
The first and the second errors are statistical and systematic (CXB fluctuations).
}
\end{tabular}
\end{center}
\end{table*}

\begin{figure}[h]
\begin{center}
\includegraphics[scale=0.36,angle=-90]{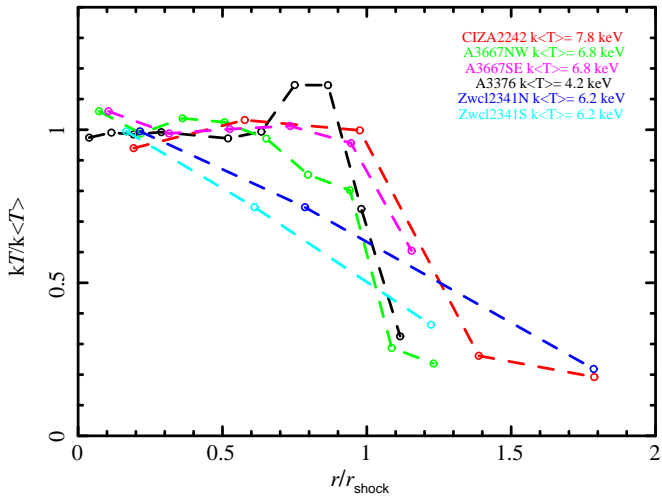}
\end{center}
\caption{
Scaled projected temperature profiles. 
The profiles are normalized by the mean temperature and
the radius of radio relic $r_{shock}$.
The $r_{shock}$ is derived from literatures summarized in Table~\ref{tab:target}. }
\label{fig:scaledtemp}
\end{figure}

\subsection{The scaled temperature profiles}\label{sec:scaledtemp}
We normalized the radial temperature profile of each cluster by the flux-weighted average temperature.
The resulting scaled temperature profiles are shown in Fig~\ref{fig:scaledtemp}.
We find that the temperature profiles of four relics, 
including the previous two results, exhibit significant discontinuities across the radio relic. 
These discontinuities can be associated with the shock front found in the radio relics.
For the other two samples (Zwcl2341N,S),  we could not see any jump in temperature profiles.
The angular resolution of Suzaku was not enough to check the presence of a discontinuity. 
Therefore, we focus on these four examples of temperature discontinuity.

Another remarkable point of the scaled temperature profile is its flatness within the radio relic.
Recent X-ray results show that the ICM in relaxed clusters has the "universal" decline temperature profile up to virial radius~\citep{markevitch98, degrandi02, vikhlinin05,pratt07, akamatsu11}. The flatness of temperature within the radio relic might need a heating mechanism, such as shock heating.

As mentioned above, the temperature of relaxed clusters tends to have the "universal" profile. It is natural that the flat temperature profiles of merging cluster will settle to the universal profile in less than a Hubble time. 
 In our samples, the locations of the radio relic are far from cluster center, typically $r_{shock} \sim$ Mpc ($0.5r_{200}$). Due to the low density ($n_e=10^{3-5}~\rm cm^{-3}$) in the outskirts of clusters, 
the cooling time in those regions exceeds the Hubble time. We might consider other processes to achieve the universal profile.

\begin{figure}[]
\begin{center}
\includegraphics[angle=-90,width=1.\hsize]{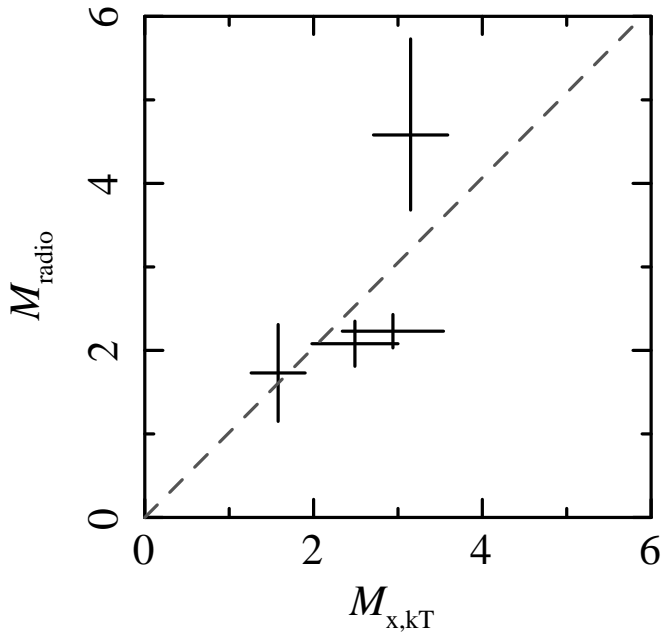}
\end{center}
\caption{
The Mach number derived from radio (${\cal M}_{radio}$) plotted against that from
the ICM temperature (${\cal M}_{X,kT}$).
The gray dashed lines indicates linear correlation as a reference.
The error bar (statistical) of  ${\cal M}_{X,kT}$ is 68 \% confidence level.  
}
\label{fig:Mach}
\end{figure}

\subsection{Mach number at Discontinuity\label{sec:Mach}}
We estimate the Mach numbers using temperature jumps across the relics. 
The Mach number can be obtained by applying the Rankine-Hugoniot jump condition, 
assuming the ratio of specific heats as $\gamma=5/3$, as
\begin{eqnarray}
\label{eq:tm}
\frac{T_2}{T_1}&=&\frac{5{\cal M}_{X,kT}^4+14{\cal} M_{X,kT}^2-3}{16{\cal M}_{X,kT}^2}, \\
\end{eqnarray}
$T_{1},T_{2}$ are the pre-shock and post-shock temperatures and ${\cal M}_{X,kT}$ is the estimated 
Mach number.  
The post and pre quantities we use are marked by red and blue horizontal bars in Figure~\ref{fig:kt}.


We compare our results with the Mach number derived from previous radio observations.
Based on Diffusive Shock Acceleration (DSA) theory, 
we estimate the Mach number based on the radio spectrum, as
\begin{equation}
\alpha=-\frac{3{\cal M}_\mathrm{radio}^{-2}+1}{2-2{\cal M}_\mathrm{radio}^{-2}},
\end{equation}
where  $\alpha$ is radio spectral index and ${\cal M}_\mathrm{radio}$ is the Mach number from the radio observation. For strong shocks, the radio spectral index will be flat about -0.5. We used radio spectral indices from the radio observations summarized in Table~\ref{tab:target}. 
The resulting Mach numbers are shown in Table~\ref{tab:mach}.
 Figure \ref{fig:Mach} displays the Mach number from the X-ray temperature and radio observations. 
 As shown in this figure, we find that the two different methods give consistent results within the current observational noise.

If we assume in the case of CIZA2242 that the mach number derived from the radio observation is correct, 
the expected post shock temperature would reach k$T=15$ keV. 
Although our results are well reproduced by a 1 temperature model,
we also fit 2 temperature models to the spectrum extracted from the \timeform{5.0'}$-~\timeform{7.7'}$ annulus,
which is located in the region where the shock just passed.
The metal abundances of the two apec components were tied to be the same, 
and only the temperature and normalization were set free.
We then obtained a statistically acceptable fit with $\chi^2$/ d.o.f.\ = 228/244
compared with the single temperature case of 231/246. 
The rather low statics of the spectrum hampered us to distinguish between the $2 kT$ and $1 kT$ models.
 The hot component temperature was derived as $kT_{\rm high}=11.5_{-3.26}^{+4.65}$~keV, and the cool one as $kT_{\rm
  low}=2.05_{-0.51}^{+1.53}$~keV.
The intensity of the cool component is about a order lower than hot one.
Based on those analyses, there were no signatures of the presence of high-temperature components.

\section{Summary}
From simulations of the structure formation of the Universe~\citep{enblin02, vazza09},
large accretion shocks and low merger shocks around clusters of galaxies are expected.
To understand the physics of cluster evolution, cluster merger events are important process.
Suzaku XIS is best suited instrument for measuring the ICM emission beyond the radio relics thanks to
its low and stable detector background.
In this paper, we analyse six radio relics of four clusters with Suzaku XIS and measured their temperature profiles. 
We found clear temperature jumps across four relics and derived the Mach number from the pre- and post temperature. The derived Mach number is almost consistent with that from the radio observation. These results strongly suggest that a shock structure exists at the radio relic.

\bigskip 
We are grateful to Takaya Ohashi and Jelle de Plaaa for insightful discussions and comments
and Ruta Kale for kindly providing  the radio index data of A3376. 
The authors thank the anonymous referee for insightful comments and suggestions which improved the content of the paper.
The authors thank the Suzaku team members for their support of the Suzaku project.  
H. A. and H.K. are supported by a Grant-in-Aid for Japan Society for the Promotion of Science (JSPS) Fellows 
(22$\cdot$1582 and 22$\cdot$5467).

\appendix
\section{Surface brightness jump in CIZA2242}\label{sec:sb_fit}
Because of the non-detection of surface brightness jump in CIZA2242,
we performed an additional check. Based on previous works~(e.g.,~\cite{2002ApJ...567L..27M}),  
we modeled the electron density profile
switching at radio relic ($r=\timeform{6.5'}$) from the beta model to a power law.  
In this model, we assume that two cases of the jump amplitude,
corresponding to the Mach number ${\cal M}=2.0$  (our
lower limit including the systematic uncertainty, see Table 4) and {\cal M}=4.5~(the radio value).
Assuming  spherical symmetry, we calculated the square 
of density as surface brightness and convolved it with the Suzaku PSF. 
The profile was binned with the same size as the observed surface
brightness.  Because the jump in the surface brightness is important, 
we selected the profile across the radio relic ($r=\timeform{5.0'}-\timeform{10.3'}$). 
Finally, we normalized the profile to the observed value at the
post  shock region ($r=\timeform{5.0'}-\timeform{7.7'} $).  The resulting profiles are 
shown in Fig.~\ref{fig:sb_fit}b (Blue and Black). 
The observed brightness profile, shown by red crosses, is consistent with
the interval of M obtained from our temperature measurements.

For the CIZA2242 case, the PSF of Suzaku corresponds to 380 kpc, which is probably
much larger than the length of shock. Hence the surface brightness jump should be significantly
diluted by other area in the bin. We need more observations with a higher angular resolution to confirm
the shock structure in the surface brightness profile.

\begin{figure*}[]
\begin{tabular}{cc}
\begin{minipage}{0.5\hsize}
(a)Modeled density profile
\begin{center}
\includegraphics[width=.8\hsize,angle=-90]{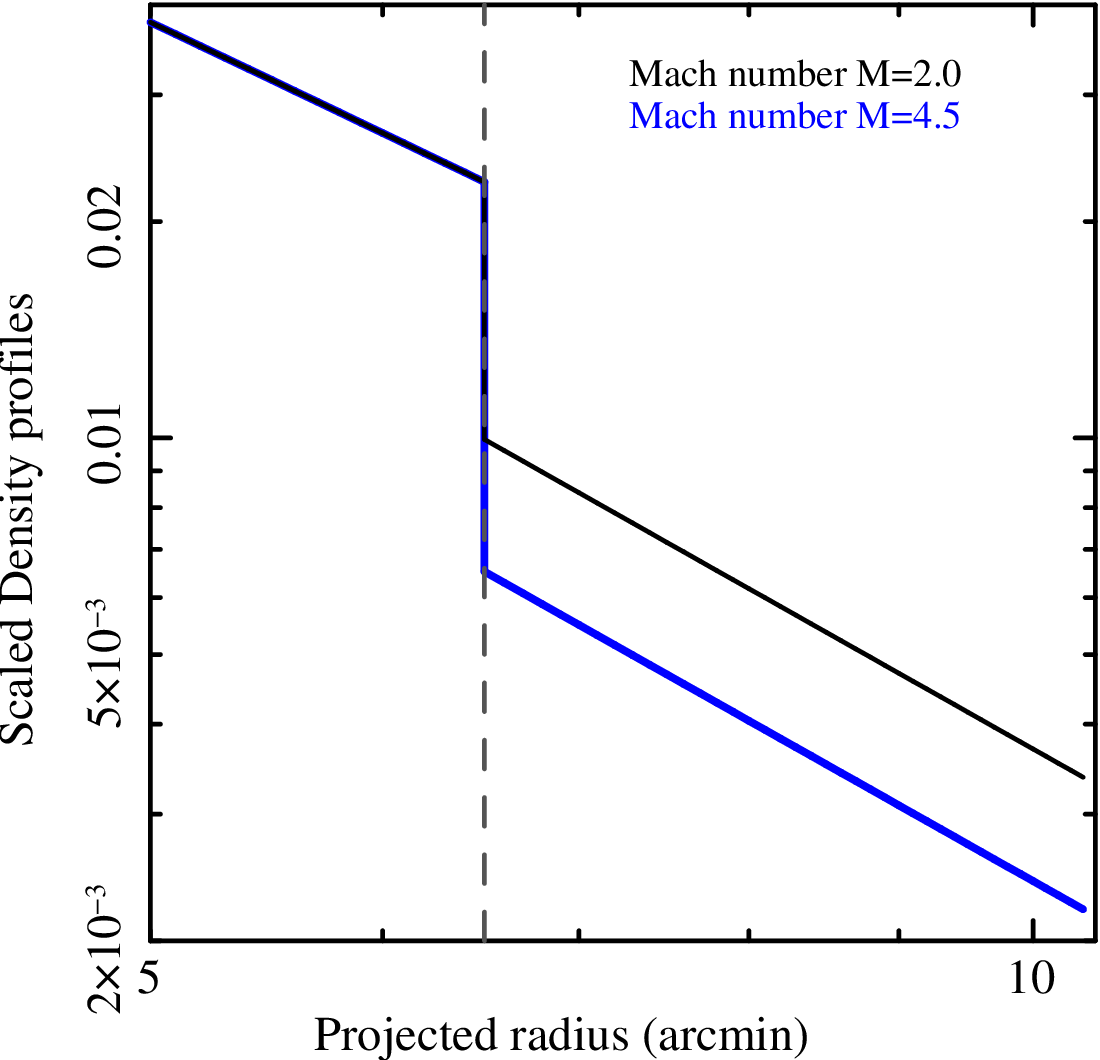}
\end{center}
\end{minipage}
\begin{minipage}{0.5\hsize}
(b)Modeled surface brightness profile
\begin{center}
\includegraphics[width=.8\hsize,angle=-90]{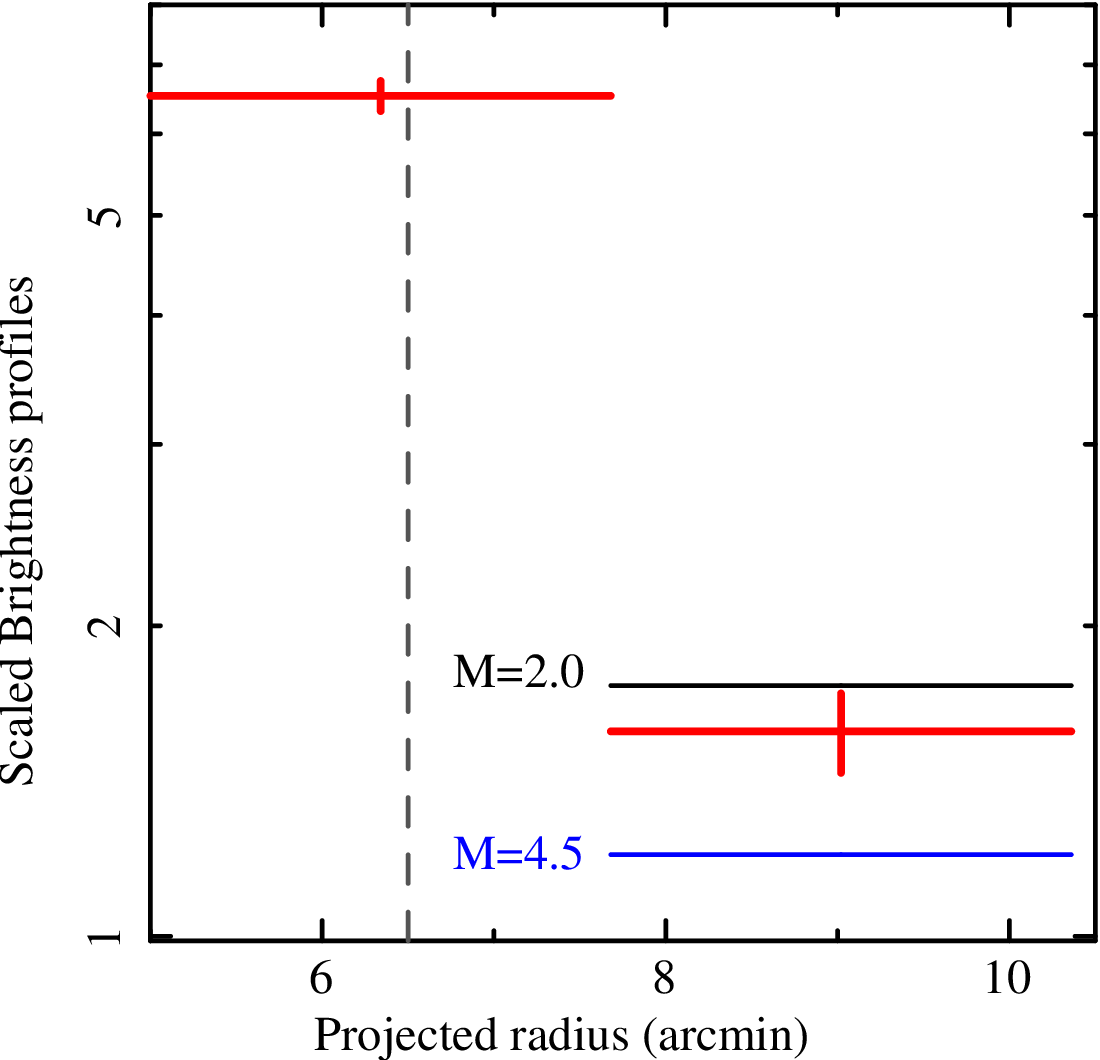}
\end{center}
\end{minipage}\\
\end{tabular}
\caption{\label{fig:sb_fit}
(a) Model used to calculate the surface brightness profile. Vertical dashed lines indicate the position the of the radio relic.  
The profile assuming the Mach number ${\cal M}=2.0, 4.5$ are shown with a black and blue lines,
respectively. The models are normalized to 1 at the center. 
(b) Surface brightness profile of CIZA2242 across the radio relic.
The red crosses show the observed value.
The expected value are shown by black (${\cal M}=2.0$) and blue (${\cal M}=4.5$) lines.
}
\end{figure*}


\end{document}